\documentclass[12pt]{iopart}

\usepackage{graphicx}% Include figure files
\usepackage{dcolumn}% Align table columns on decimal point
\usepackage{bm}% bold math

\newcommand{\edm}{\eta}
\newcommand{\edmi}{\eta_i}
\newcommand{\ledmi}{\log\edmi}

\newcommand{\edmo}{\eta_{\rm obs}}
\newcommand{\lpdmi}{{\cal P}_i} 
\newcommand{\dpdmi}{\lpdmi}
\newcommand{\physt}{${\cal T}$}
\newcommand{\oo}{{\alpha}}
\def\agt{ \lower .75ex \hbox{$\sim$} \llap{\raise .27ex \hbox{$>$}} }
\def\alt{ \lower .75ex \hbox{$\sim$} \llap{\raise .27ex \hbox{$<$}} }

\def\ie{{\frenchspacing\it i.e.}}
\def\eg{{\frenchspacing\it e.g.}}
\def\etc{{\frenchspacing\it etc.}}

\begin{document}
	
\title[Multiple universes and other dark matters]{Multiple universes, cosmic coincidences, and other dark matters}

\author{Anthony Aguirre} \address{Department of Physics, UC Santa
Cruz, Santa Cruz, CA 95064}
\ead{aguirre@scipp.ucsc.edu}

\author{Max Tegmark}
\address{Department of Physics and Astronomy, University of Pennsylvania, Philadelphia, PA 19104}
\address{Department of Physics, Massachusetts Institute of Technology, Cambridge, MA 02139}
\ead{tegmark@mit.edu}

\date{\today}

\setcounter{footnote}{0}

\begin{abstract}
Even when completely and consistently formulated, a fundamental theory
of physics and cosmological boundary conditions may not give
unambiguous and unique predictions for the universe we observe; indeed
inflation, string/M theory, and quantum cosmology all arguably suggest
that we can observe only one member of an ensemble with diverse
properties.  How, then, can such theories be tested?  It has been
variously asserted that in a future measurement we should observe the
{\em a priori} most probable set of predicted properties (the
``bottom-up'' approach), or the most probable set compatible with all
current observations (the ``top-down'' approach), or the most probable
set consistent with the existence of observers (the ``anthropic''
approach).  These inhabit a spectrum of levels of conditionalization
and can lead to qualitatively different predictions.  For example, in
a context in which the densities of various species of dark matter
vary among members of an ensemble of otherwise similar regions, from
the top-down or anthropic viewpoints -- but not the bottom-up -- it
would be natural for us to observe multiple types of dark matter with
similar contributions to the observed dark matter density.  In the
anthropic approach it is also possible in principle to strengthen this
argument and the limit the number of likely dark matter
sub-components. In both cases the argument may be extendible to dark
energy or primordial density perturbations.  This implies that the
anthropic approach to cosmology, introduced in part to explain
``coincidences" between unrelated constituents of our universe,
predicts that more, as-yet-unobserved coincidences should come to
light.

\end{abstract}

%\submitto{\JCAP}


\maketitle

\section{Introduction}
\label{sec-intro}

Our surrounding physical universe is very well described by general
relativity and the (quantum mechanical) standard model of particle
physics, combined with initial conditions specifying a hot big-bang
cosmology dominated by dark matter and dark energy.  But this
description is fundamentally incomplete: explanations for dark matter
and dark energy lie outside of the standard model, general relativity
is not a quantum theory, and general relativity and the standard model
cannot address the big-bang singularity or initial conditions.

The current most highly regarded candidate theories of more
fundamental physics and of big-bang initial conditions appear to
be, respectively, string/M theory and inflation.  Although it is not
clear how (or whether) these theories may be combined into some
consistent quantum cosmology, it is interesting---and somewhat
disquieting---that string/M theory, inflation, and current
formulations of quantum cosmology {\em all} share an important
feature: it is far from clear that any of the three components (even
less their potential combination) makes a single, unique set of
predictions regarding our observable world.

In current formulations of string/M theory, low-energy physics -- such
the cosmological constant $\Lambda$, the particle physics coupling
constants, and the cosmological parameters -- is governed by dynamical
degrees of freedom called {\em moduli} that are in turn governed by a
low-energy effective potential which is itself determined by a set of
field fluxes. Metastable minima of this potential constitute vacua,
one of which would determine the physics of our world.  There appears
to be no unique vacuum (there is a continuous space of different
supersymmetric $\Lambda=0$ vacua), nor any good reason to suppose that
there is just one (or even small number of) non-supersymmetric
$\Lambda > 0$ vacua such as could describe our universe~(see,
\eg~\cite{Susskind:2003kw,Douglas:2003um,Kachru:2003aw,Kachru:2003sx,Douglas:2004zg}
but also~\cite{Banks:2003es}).

In generic models for inflation, the observable universe is but one of
many thermalized regions spawned within an eternally inflating
background~\cite{Guth:2000ka,Linde:2002gj,Vilenkin:1999pi}. Due to the
exponential expansion, each region may be described by a Friedmann
cosmology with some governing parameters, but fields governing these
parameters (including the M-theory moduli) may be globally
inhomogeneous so that the parameters vary from one region to another.
Thus such inflation can lead to an ensemble of ``sub-universes'' in
which both particle physics coupling constants and cosmological
parameters such as the amplitude of primordial
fluctuations~\cite{Vilenkin:1999pi}, the cosmological
constant~\cite{Garriga:1999bf,Kallosh:2002gg} or the density of dark
matter~\cite{Linde:1987bx} may vary.

In formulations of quantum cosmology such as the ``no boundary
proposal'' using Euclidean quantum gravity, a prescription is given
for computing the amplitude of a given spatial section and field
configuration.  These are then pieced together to give amplitudes for
cosmological histories.  In the ``many worlds'' interpretation of
quantum mechanics\cite{Everett}, all of these histories are equally real and
co-exist in superposition.  Cosmological parameters such as the
curvature scale~\cite{Hawking:2002af,Hawking:1998bn} or
others~\cite{Turok:1998he} can then take a random value in each world,
drawn from a fairly well-defined probability distribution.

If we are fortunate, these ways of understanding of quantum gravity,
inflation, and quantum cosmology will all turn out to be in some
essential way incorrect or inapplicable to a true ``fundamental
theory'' of cosmology that will instead give unique predictions.  But
if not, and at least in the meantime, we must face the rather thorny
issue of how to extract meaningful predictions from a theory that
predicts a ``multiverse'', by which we mean an ensemble of
physically-realized systems with different properties, only one of
which may possibly be observed by any given observer. This is the
subject of the present paper.

Three approaches to this problem have been widely adopted: the
``bottom-up'', ``top-down'', and ``anthropic'' approaches. The meaning
of these terms will become clear in Sec~\ref{sec-methods}.  This paper
argues for three basic points.  First, that in a cosmological context
with no {\em unique} predictions, these different approaches are of
more than academic or philosophical interest: they correspond to
different specific questions being implicitly asked, and can lead to
genuinely different answers to the more general question of ``what
will we observe?''  Second, that the top-down and bottom-up
approaches should be seen as two ends of a spectrum, with anthropic
arguments in-between. Third, that when applied to the specific subject
of dark matter, what we expect to see in future observations depends
qualitatively on on which method of making predictions is adopted, and
that in the top-down or anthropic approaches a rather
counter-intuitive general prediction emerges.

The plan of the paper is as follows.  In Sec.~\ref{sec-methods} we
discuss the three basic ways of making predictions, and how they may
be viewed in a unified way.  In Sec.~\ref{sec-applic} we make generic
predictions regarding dark matter in three different ways, in the
context where there are many possible types of dark matter with
densities that vary from one member to another of an ensemble of
regions that otherwise resemble our observable universe.  The
extension of the arguments to other cosmological quantities is
discussed in Sec.~\ref{sec-per}, and the general issue of ``cosmic
coincidences" is analyzed in Sec.~\ref{sec-coinc}. We draw general
conclusions in Sec.~\ref{sec-conc}.

\section{The spectrum of conditionalization}
\label{sec-methods}

Imagine that we have a physical theory and set of cosmological
boundary conditions (or ``wave function of the universe"), \physt, that
predicts an ensemble of physically realized systems, each of which is
approximately homogeneous in some coordinates and can be characterized
by a set of parameters that may vary from one system to another.
Denote each such system a ``universe'' and the ensemble a
``multiverse''.  Given that we can observe only one of these
universes, what conclusions can we draw regarding the correctness of
\physt, and how?  This is the problem apparently confronting -- at
varying levels -- eternal inflation, quantum cosmology in the
many-worlds interpretation, and string/M theory.

%[[can I fix footnotes?]]
Because the alternative is to give up at the outset, let us assume
that we have {\em some} quantitative way of comparing these
universes\footnote{For this reason we will not consider here truly
disjoint multiverses with no unifying physics \physt;
see~\cite{Tegmark98,Ellis:2003dx} for a discussion of such multiverses
and the rather daunting difficulties associated with them.} so that we
may specify the joint probability distribution ${P}(\oo_1,...,\oo_N)$
for observables $\oo_i$ ($i=1..N$). To see what this might involve,
consider the combination of the string/M theory landscape with eternal
inflation.  Let $\alpha_i$ be low-energy particle physics or
cosmological parameters that may be compared to observations. To
realize predictions of these $\alpha_i$ we might use the following
successively definable quantities:
\begin{enumerate}
\item The relative number $N_{\rm vacua}(\alpha_i,\Delta\alpha_i)$ of
vacua with low-energy constants in the range
$[\alpha_i,\alpha_i+\Delta\alpha_i]$ for each $i$.
\item The relative number $N_{\rm real}$ of these vacua that actually
come into being in the cosmological context; this may differ
exponentially from $N_{\rm vacua}$ because of exponentially
suppressed tunneling (or transition) rates between vacua. 
\item The relative number $N_{V}$ of physical planck volumes on
the post-inflation reheating surface in all FRW cosmologies for which
the range of low-energy constants holds. This can easily differ
exponentially from $N_{\rm vacua}$ due to different periods of
inflation (and may also involve ambiguous comparisons between infinite
volumes\cite{Linde:1995uf,Vanchurin:1999iv,Guth:2000ka,Garriga:2001ri,Tegmark:2004qd}.)
\item The relative number $N_B$ of baryons that exist in these
volumes.  These may be quite different from $N_V$ depending
on the physics of baryogenesis.
\end{enumerate}

None of these quantities promises to be easy to calculate, and each
is harder than the previous one.  But without calculating at least
one, the theory makes no predictions whatsoever. For present purposes
we will call $P(\alpha_i)$ {\em a priori} probabilities which are
calculated using one of $N_{\rm vacua}, N_{\rm real},N_{V}$ or $N_B$.
For example, we may use $N_B$ to calculate ${P}_B(\oo)$, the
probability per unit $\oo$ that a randomly chosen baryon would inhabit
a universe in which $\oo$ takes the given value. (For instance, 
suppose that only one observable $\oo$ varies among
the universes, that each universe has a finite baryon
number $B$ depending only on $\oo$, and that each universe has a value of
$\oo$ drawn at random from some cumulative probability distribution
$p(\oo)$; then ${P}_B(\oo)\propto B(\oo)dp(\oo)/d\oo$
would be the probability per unit $\oo$ that a randomly chosen baryon would
inhabit a universe in which $\oo$ takes the given value.) 

With ${P}(\oo_i)$ in hand, we wish to connect our observation of the
$\oo_i$ to \physt; but how?  The first approach one might consider, which
may be termed the {\it ``bottom-up''} approach\footnote{The idea of
this nomenclature is that \physt, being fundamental, provides the
foundation for a logical structure -- of which the succession of quantities
$N_{\rm vacua}, N_{\rm real},N_{V},$ and $N_B$ would be part --
with our particular low-energy,
local observations at the top.}, is to use ${P}(\oo_i)$ as directly as
possible and assert, for example, that for each $k$ we should observe
$\oo_k$ to be near the peak of the probability distribution obtained
by marginalizing ${P}(\oo_i)$ over $i\neq k$. If not, then \physt\ is
ruled out at some confidence level depending upon the shape of
${P}(\oo_i)$ and the observed value of $\oo_k$.  This seems
straightforward, but hides an unavoidable choice that was made between
weighting by baryons (using $P_B$) rather by volume (using $P_V$), or
giving each universe equal weight (\ie, using $P_{\rm vacua}$ or
$P_{\rm real}$).  Implicitly we are asking: ``What sort of universe
should I live in, given that I am a randomly chosen baryon?'', rather
than ``What should I observe given that I am a random chosen volume
element?''  or ``Given that I am a randomly chosen universe, what type
of universe should I be?'' It is rather unclear which, if any, of
these questions bears upon an observation that we make.  Indeed, the
only completely unambiguous situation would seem to be that in which
$\oo$ is completely independent of any other physical property and so
is simply a random variable that attains, in each universe, a value
drawn from the distribution ${P}(\oo)$.  But none of the cosmological
parameters are of this nature.

A second, quite different approach -- which may be called {\it
``top-down''} -- asks the question differently: ``given everything
thus far measured, along with \physt, what should I measure for
$\oo_k$?'' In making a prediction, one thus conditions on all
available data (including $\oo_i$ for $i\neq k$), implicitly
discarding from consideration all regions with properties different
than those observed in our region.  The danger of this approach is
that it leads to the acceptance of proposals for \physt\ for which the
properties of our observed universe are incredibly rare.  Outside of
cosmology, this may not be worrisome: in order to test a theory we may
construct a very special experimental arrangement and not worry that
is ``improbable''.  But in cosmology there is no wider context, and
the acceptance would go directly against normal scientific
methodology: if a theory predicts value $a_1$ for some observable $A$
with 99.99999\% probability, and result $a_2$ with 0.00001\%, one
would be reluctant to accept the theory if a single trial were
performed and result $a_2$ were obtained; why should this change if
$A$ has been measured previously and one is now measuring observable
$B$?  For the top-down approach to make sense, it seems, one must
implicitly posit that there is some {\em reason} to neglect the
discarded regions, so that the conditionalization is justified.

The third approach, the {\it ``anthropic''}, attempts to remedy the
shortcomings of the others by supplying the appropriate weighting
quantity lacked by the bottom-up approach, and at the same time
providing the reason required by the top-down approach for discarding
regions unlike our own.  The anthropic approach innocuously enough
attempts to derive the probability distribution of {\em observed}
parameter values, and thus implicitly asks: ``given that I am a
typical observer, what value of $\oo$ should I expect to measure?''
(The key assumption, that we should expect to observe the same value as a
typical observer, been called the ``principle of
mediocrity''~\cite{Vilenkin:1994ua}.)  The anthropic approach is,
therefore, the bottom-up approach weighted by ``observers'', and at
the same time an attempt to justify the top-down approach by asserting
that regions very different from our own have few or no observers in
them.  The clear problem of
principle\footnote{See~\cite{Aguirre:2001,Ellis:2003dx} for some
discussion of the great technical problems.} in the anthropic approach
is in how to define ``observer.''  One might ask, for example, ``given
that I am a living being, what value of the cosmological constant
$\Lambda$ should I observe'' and obtain indeed a prediction for
${P}(\Lambda)$ but the concomitant prediction that one is most likely
an insect or bacterium.  Or it might be asked ``given that I am in a
galaxy, what is ${P}(\Lambda)$?'' But the results would then depend on
the galaxy mass chosen, and may be wrong if galaxies are either not
required, or not all that is required, for the existence of observers.

The bottom-up and top-down approaches form the ends of a spectrum of
conditionalization, with the anthropic approach (or any other approach
that conditions on some but not all available observations) in
between.\footnote{We note that Hartle~\cite{Hartle:2004qv} has
outlined the problem in quantum cosmology in somewhat similar terms,
and Bostrum's notion of a ``reference class"~\cite{bostrum} is also
similar.}

  The bottom-up approach has the maximal power to rule out proposals
for \physt, but may rule out the correct one.  The top-down approach is
sure to allow the correct theory, but may additionally allow many
other erroneous ones.  The desirable compromise (which seems to be the
implicit hope of proponents of the anthropic principle) would appear
to be to condition on as little as possible, while still making
entirely accurate predictions of all data not conditioned on.  But it
is unclear how we can know when minimal necessary conditionalization
has been reached.  This is the heart of the conundrum.

One might hope that the distinctions between top-down, bottom-up,
{\etc} would be academic when it comes to making real predictions about
the real universe. There is, however, no reason to believe that this
is the case.  Clearly different conditionalizations will lead
to different probabalistic predictions from the same candidate \physt.
But just as clearly, this will lead to different inferences of the
types of \physt s capable of matching observations -- a \physt\ that would
reproduce our observations only by astonishing luck or coincidence
(and hence which we would be apt to discard) under one
conditionalization might seem very natural under another.  Thus what
may ``naturally" be observed in future observations may differ, in a
rather general way, upon which type of reasoning (\ie, what
conditionalization level) is used.

In the next section, we more explicitly argue that this is 
the case, using an extended and explicit example of predictions
for dark matter.

\section{Application to dark matter}
\label{sec-applic}

\subsection{The context}

Consider, as discussed above, a proposal for \physt\ that predicts an
ensemble of ``universes'', each of which may be accurately modeled as
a big-bang described by $N$ cosmological parameters with values
$\alpha_k (k=1..N)$.  Now suppose, contrary to convention, that there
are $N_{\rm DM}> 1$ different {\em independent} substances that act as
collisionless, nonbaryonic dark matter, all of which exist in the
low-energy phenomenology of \physt.  Many such particles exist in the
literature, {\eg} neutralinos and other supersymmetric particles,
axions, sterile neutrinos, Kaluza-Klein particles, cryptons, light
scalar particles, ``little Higgs" particles, Q-balls, monopoles,
WIMPzillas, LIMPs, CHAMPS, D-matter, Brane-world dark matter, mirror
matter, quark nuggets, primordial black holes, \etc, {\it etc.}  Not
{\em all} dark matter candidates in the literature can even in
principle coexist -- for example there can be only one lightest
supersymmetric particle in each universe. But let us assume that the
phenomenology of \physt\ is rich enough that some number $N_{\rm DM}$
of them are present in our full ensemble, with the density of each
species $i$ in each universe characterized by a dark-matter-to-baryon
ratio $\edmi \equiv \Omega_{\rm DM,i}/\Omega_b$, which may be zero.
Examples of ways in which densities for a number of dark matter
candidates may become probabalistic are given in~\cite{Hodges:1991zf}.

For tractability, let us assume that each is governed by an {\em
independent} normalizeable cumulative probability distribution
$p_i(\edmi)$ that describes the probability that a randomly chosen
baryon resides in a universe with the given $\edmi$. Of course in
reality the probabilities may well not be independent, nor independent
of the $\alpha_k$; but the present purpose is to explore differences
in predictions for a {\em given} $p_i[\edmi]$, and independent
probabilities are the simplest case.  For convenience we will
generally work with the differential distribution in $\log\edmi$,
$\lpdmi\equiv dp(\edmi)/d\ln\edmi= \edmi dp(\edmi)/d\edmi$. Let the
remaining $N-N_{\rm DM}$ parameters $\alpha_k$ be described by a
similarly defined probability distribution
$p(\alpha_1,...\alpha_{N-N_{\rm DM}})$.  These probability
distributions may also describe parameters taking discrete values by
composing $p_i(\edmi)$ and $p(\alpha_1,...\alpha_{N-N_{\rm DM}})$ of
$\delta$-functions, but we will assume that the distributions are
smooth, or that the allowed discrete values of $\edmi$ are closely
enough spaced to be described by a smooth distribution.

\subsection{Argument using bottom-up/minimal conditionalization}

Let us now ask, using the bottom-up approach, what we expect to
find in future experiments bearing upon the nature of the dark matter.
First, we must suppose that our \physt\ proposal predicts that the
probability distribution $P(\alpha_1,...\alpha_{N-N_{\rm DM}})$ for
{\em already measured} parameters $\alpha_k$ is such that those values
are reasonably probable, or even uniquely and correctly predicted --
otherwise we should already have discarded \physt.  (We are thus
assuming, here and in subsequent bottom-up reasoning, that the
parameters we observe are not highly unlikely given \physt.)

Now, each $\edmi$ might in principle take any order of magnitude; for
definiteness, let us say that each $\ledmi$ has a range of $M\gg
1$.\footnote{Of course, a specific \physt\ will only allow a specific
range for each $\ledmi$, but this is taken into account by the
$\lpdmi$ which would be nonzero only over this range.}  It would then
require fine-tuning for any given universe two dark matter components
would have $\edm$ of the same order. This is because each component
will have a $\lpdmi$ that is either large only across a narrow range
of $\log\edm$ (in which case it is unlikely for that range to overlap
with the $\edm$ of the second component), or will have a $\lpdmi$ that
is large over a very wide range of $\log\edm$ (in which case it is
unlikely that in the given universe the $\edm$ chosen from this wide
distribution will agree with the second component).
Very roughly, we might estimate the degree of fine-tuning by
assuming that the ${\cal P}_i$ peak at uniformly random values of
$\log\edm$, and thus that the probability $P$ of the $j$
highest-density components having the same order of magnitude would
just be given by the binomial distribution:
\begin{equation}
P=p^{(j-1)}(1-p)^{N_{\rm DM}-j}{ (N_{\rm DM}-1)!\over (N_{\rm DM}-j)!(j-1)!},
\label{eq-probcoinc}
\end{equation}
where $p=1/M$, and the average number $\bar j$ of components
coincident with the dominant one will simply be $\bar j=(N_{\rm
DM}-1)/M$.  Both numbers will be very small unless $N_{\rm DM} \agt M$,
{\ie} $N_{\rm DM}$ is extremely large.

By this argument, in the absence of any genuine information on the
$\lpdmi$, the only likely occurence from the bottom-up approach is to
have the dark matter dominated by one species, with all other possibly
existent species giving negligible contributions.  If the dominant
species is predicted to have $\edm\approx \edmo \simeq 5$ (the
observed value) then the \physt\ proposal may be accepted; otherwise it
must be discarded and the argument repeated with some new proposal for
\physt.

The preceding elaborate argument has thus arrived at the standard
conclusion that given several dark matter types with independent
physics (\ie, in the absence of any physical reason why the $\dpdmi$
would peak narrowly near the same values), we would be surprised if
they all contributed similarly to $\Omega_{\rm DM}$, for just the same
reason it is generally considered suprising that the energy densities
of dark matter, baryons, neutrinos and dark energy are all presently
comparable.  

Note that this is not, of course, a deductive prediction that could be
obtained from any given \physt.  Rather, it is in inference drawn by
comparing many imagined prospective \physt-candidates to a set of
observations. This is something we do routinely: for example, in
``deriving" cosmological parameters from astronomical observations we
implicitly have in mind an imaginary ensemble of universes which we
can compare in turn with our observations to select out the set that
is in reasonable accord with those observations.  In doing that
analysis, bottom-up reasoning weighted by universe is generally used,
which seems sensible given that the ensemble is imaginary; but the
weighting could be done differently (\ie, different ``priors" could be
chosen, or alternatively a different measure could be placed over the
space of theories) and different inferences would be so obtained.  In
general we know of no reasonable way to define such a measure over
``theory space'' and we made a simple assumption in calculating
Eq.~\ref{eq-probcoinc}; however we note that this is also implicitly
done in {\em any} discussion of whether a given theory is fine-tuned.
The present argument is of just this character.

\subsection{Argument using top-down/maximal conditionalization}

Let us now perform the analysis from the top-down point of view.  We
again assume that there is an ensemble of regions with $N_{\rm DM}$
species of dark matter of densities $\edmi$ present in each.  Now,
however, we further assume that we inhabit one member of a
sub-ensemble of universes in which the $\alpha_i$ all take values
compatible with current observations; in particular we assume that
whatever values the individual $\edmi$ take, the total dark matter
density $\edm\equiv\sum_i\edmi$ satisfies $\edm=\edmo$, since this is
essentially all that is known about the dark
matter.\footnote{Implicitly, we are assuming that all of the
components genuinely are cold, weakly interacting, particles that
could not have been detected by current experiments. Stronger
constraints could be imposed on, \eg, warm or hot dark matter.}  This
sub-ensemble may constitute a tiny part of the full ensemble, but in
the top-down approach we refuse to be troubled by this.

	While we know $\edm$, we do not know the individual $\edmi$,
so we can predict their most probable values using our candidate \physt.
This requires that we maximize the total probability
\begin{equation}
{\cal P}_{\rm tot} \propto \prod_i\lpdmi(\ledmi),
\label{eq:sumprob}
\end{equation}
subject to $\edm=\edmo.$ Each $\lpdmi$ may favor (\ie\ have
 substantial probability at) ``low'' values $\edmi \ll \edmo$, ``high
 values'' $\edmi \gg \edmo$, the observed value $\edmi \sim \edmo$, or
 some combination of these.  Now pick some component $i$. We have
 argued above that it is {\em a priori} unlikely that only
 $\edmi\sim\edmo$ will be likely.  Then, if low values are favored
 (and whether or not high or observed values are), then they are
 probably attained, since there is no conflict with the
 conditionalization $\edm\simeq \edmo$; therefore each such component
 would be expected to have an unobservably small density in our
 universe.

This leaves us with $N_{\rm big}$ components with probable values near
or far above $\edmo$.  Let us then examine the set of $\dpdmi$ for
these near $\edmi\sim \edmo$, where they may have small (but
nonzero\footnote{If for any of the components favoring high $\eta_i$,
{\em no} universes have $\ledmi<\log\edmo$, then of course \physt\ is
ruled out; this would be wonderful as it would be a definite
prediction of \physt.})  probability.

We can compute the maximum of the total probability ${\cal P}_{\rm
tot}$ subject to the condition of fixed $\edm=\edmo$ using a Lagrange multiplier:
\[     \max \left\{ \ln {\cal P}_{\rm tot}  - \lambda \sum_i \eta_i\right\}\]
gives
\[    \max \sum_i [\ln {\cal P}_i - \lambda {\eta_i}]\]
and the solution
\begin{equation}   
d\ln {\cal P}_i/d\eta_i  = \lambda.
\label{eq-gensol}
\end{equation}
In other words, the optimal values $\eta_i$ have the intuitive property 
that increasing any of them by a small amount increases $\ln {\cal P}_{\rm tot}$ by the same factor $\lambda$.

For a more explicit solution we will assume that each is locally a
power-law, \ie
\begin{equation}
\dpdmi \propto \edmi^{\beta_i}.
\label{eq:plansatz}
\end{equation}
Since we are considering a narrow range of $\edm\sim\edmo$, this
should be a good approximation unless there is a preferred scale in
the probability distribution of order $\edmo$, which is {\em a priori}
unlikely (i.e. would require fine-tuning).  We also assume $\beta_i >
0$ as we are only considering the $N_{\rm big}$ components with
probabilities peaked at high $\eta$.

The maximization of ${\cal P}_{\rm tot}$ subject to
$\edm=\edmo$ yields, using Eq.~\ref{eq-gensol},
\begin{equation}
\edmi =\left({\beta_i\over\beta}\right)\edmo,\ \ \beta\equiv \sum_j \beta_j,
\end{equation}
thus the contributions to $\edm$ of the $N_{\rm big}$ components will
be different by orders of magnitude only if the power-laws governing
the probability distributions at $\edmi\sim \edm^{\rm obs}$ are.  This
cannot be ruled out (although this may be possible in the anthropic
approach as discussed below), but it may be argued for some specific
cases that this is unlikely; for example, in discussing the
cosmological constant in a similar context,
Weinberg~\cite{Weinberg:2000qm} (see
also~\cite{Vilenkin:1995nb,Garriga:1999bf,Smolin:2004yv,Hodges:1991zf}) argues that
because the range considered (here $\edm\sim\edmo$) is so small
compared to the characteristic scale governing $P(\edmi)$, the latter
should locally be nearly flat (\ie, $\beta=1$ here).

Three conclusions have been drawn using the top-down approach,
summarized as follows.  First, it is very likely that we live in a
very improbable ensemble member, because at least one dark matter
component $i$ is likely to have $\dpdmi$ peaked at $\edmi \gg \edmo$.
But we have decreed that we shall accept this in light of observed
facts.  Second, as in the bottom-up case, a number of components
should contribute $\edm \ll \edm^{\rm obs}$, and we shall probably not
detect these.  Third, several components should contribute roughly
equally to $\edm^{\rm obs}$, unless one of the components has a
probability distribution that is both peaked at $\edm \gg \edm^{\rm
obs}$ and is quickly varying at $\edm \sim \edmo$.  We shall call this
the ``principle of equal representation" (PER): when a
conditionalization is placed on a {\em sum} $\alpha\equiv
\sum_i\alpha_i$ of parameters with (relatively slowly) rising
independent probability distributions ${\cal P}(\alpha_i)$, the most
probable combination subject to the conditionalization is that all
$\alpha_i$ are of similar order.

Thus the generic prediction of the top-down case is directly contrary
to the result obtained in the bottom-up case; whereas in the latter we
would expect one dominant dark matter form, here it is quite
reasonable to expect several (even many) forms to have comparable
contributions.  However, we cannot eliminate the possibility of an
extremely high $\beta_i$ value that would allow one component to be
dominant, nor nor have we explained why we live in a very improbable
region.

\subsection{Anthropic argument/partial conditionalization}

The final approach we may take is that of {\em partial}
conditionalization, which includes the anthropic approach.  Starting
with the same ensemble of universes each with $N_{\rm DM}$ dark matter
components, we calculate probabilities as in the bottom-up approach,
but with an additional weighting (or ``conditionalization) factor
$W(\alpha_k)$ applied to the baryon-weighted (or volume-weighted)
probabilities.  To reproduce the top-down approach, for example, we
may set $W(\alpha_k)=0$ if $\alpha_k$ is incompatible with our
observations for any $k$, and $W=1$ otherwise. Then if the
probabilities are re-normalized, this is equivalent to limiting the
analysis to a sub-ensemble of universes compatible with our current
observations.  But $W$ can be much less restrictive. For example,
$W(\alpha_k)$ could count the number of some object ${\cal X}$ (say
a large spiral galaxy) per baryon in a universe with parameters
$\alpha_k$.

The (re-normalized) $W$-weighted probabilities can be interpreted in
two different ways.  First, they could be said to describe what we
expect to measure for $\alpha_k$ given only that we observe an ${\cal
X}$. One might then view this ``partial conditionalization'' approach
simply as relaxing the top-down assumptions, in order to gain more
predictive power using less
assumptions~\cite{Albrecht:2002uz,Hawking:2002af,Hawking:2003bf}; but
why keep some conditions, and not all, or none?

Alternatively, one might with a slight philosophical shift take the
anthropic approach of attempting to ask: ``given that I am a randomly
chosen {\em observer}, what should I observe?''  Then the ${\cal
X}$-object should be an observer, and the $W$-weighted probabilities
would describe the probability that a randomly chosen observer
inhabits a universe with parameters $\alpha_k$.  This clears up some
of the ambiguity of the partial conditionalization approach in that
there is a justification conditionalizing on observers.  However,
since there is no obvious definition of what ``observer'' actually
means, one is forced to adopt some proxy -- stars, galaxies, or
universes ``pretty much'' like ours -- and hence the ambiguity
remains.

A third possibility, often employed in the literature in making
anthropic predictions, is to focus on a single parameter, say
$\Lambda$, fix all other cosmological parameters to the observed
values, then weight ${\cal P}(\Lambda)$ by some some proxy for
observers, {\eg} by a $W(\Lambda)$ that is the number of galaxies
per baryon (or unit volume) in a universe of the given $\Lambda$. This
approach is generally taken for purposes of tractability, but is
really justified only if $\Lambda$ {\em alone} varies across the
ensemble; the predictions made by varying only one parameter will
generally {\em not} be the same as if several are varied (see
Sec.~\ref{sec-coinc} and
Refs.~\cite{Tegmark:1997in,Aguirre:2001,Pogosian:2004hd,Graesser:2004ng}),
and if several parameters vary, it seems hard to justify treating only
a subset of them anthropically.

In any case, once a conditionalization (or weighting) factor is
 chosen, a calculation can be done to predict the various dark matter
 densities by computing the total probability that an ${\cal X}$
 resides in a universe with defining parameters $\alpha_k$ and dark
 matter ratios $\edmi$:
\begin{equation}
{\cal P}_{\rm tot}={\cal P}(\alpha_k)\times\prod_i\lpdmi\times W(\alpha_k,\edm),
\label{eq:ptot}
\end{equation}
where $W$ is the number of ${\cal X}$s per baryon in a universe
with parameters $\alpha_k$ and total dark matter density
$\edm$.\footnote{We are assuming here that only the total dark matter
density is important in forming ${\cal X}$-objects.}

Now there are three possibilities.  Either 1) ${\cal P}_{\rm tot}$
 allows reasonable probability of the observed values of
$\alpha_k$ and $\edm$, and $W(\edm)$ is falling near $\edm\sim\edmo$,
or 2) the same, but $W(\edm)$ is rising at $\edm\sim\edmo$, or 3)
${\cal P}_{\rm tot}$ has little probability near the observed values.

In cases 1) and 2), we have a good candidate for an anthropic
explanation of the cosmological parameters $\alpha_k$ and $\edm$ {\em
if} we can make a strong argument that our ${\cal X}$-object is a good
proxy for an observer.  But what do we predict for the individual dark
matter densities $\edmi$?  In case 1, the situation is quite similar
to that in the top-down approach, and the same arguments lead to the
conclusion that it is natural for several dark components of
comparable density to contribute to $\edm$, with several other
components existing at undetectable levels.  Thus the PER applies to
the anthropic standpoint, {\em if} there is a high-$\edm$ cutoff in
the hospitality factor $W(\edm)$.  In case 2, however, the PER will
not hold, because ${\cal P}(\edmi)$ must be falling for all $i$ (or
else ${\cal P}_{\rm tot}$ could not fall off on both sides of $\edmo$,
and we would instead be discussing case 3).  In this case ${\cal
P}_{\rm tot}$ will be greatest when only {\em one} dark matter
component is required to take an improbable value $\edmi\sim\edmo$.

Similarly, additional inferences can in principle be made using using
case 3 that ${\cal P}_{\rm tot}$ does {\em not} peak near the
observed parameter values. In this case we are led to conclude that
the fundamental theory and its ${\cal P}$ are incorrect.  Given a form
of $W$, we can in this way constrain the form of ${\cal P}$.  To see
this more explicitly, suppose that we have a candidate \physt\ with
$N_{\rm DM}$ dark matter components for which $\lpdmi \propto
\edmi^{\beta_i}$ at $\edm \sim\edmo$, and for which all other
parameters have fixed, unique values.  Then let us maximize
\begin{equation}
{\cal P}_{\rm tot}(\edmi) \propto \prod_i \edmi^{\beta_i}W(\edm).
\end{equation}
Because the conditionalization factor $W$ depends only on the {\em
total} dark matter density $\edm$, the maximum of ${\cal P}_{\rm tot}$ will
occur where each $\edmi = \beta_i/\beta$ (where $\beta\equiv \sum_i
\beta_i$), as in the top-down approach.  But furthermore, the peak
will occur at a value $\eta_{\rm max}$ that depends upon $\beta$.  

For example, let us model $W$ (for demonstrational purposes only) by
either a power-law falloff (with index $\gamma$) or Gaussian decline
for $\eta > \eta_0$, \ie
\begin{equation}
W(\edm)\propto {1\over 1+(\eta/\eta_0)^\gamma}\quad{\rm or}\quad
W(\edm)\propto e^{-\eta^2/2\eta_0^2}.
\end{equation}
If ${\cal
P}(\eta_i)\propto \eta_i^{\beta_i}$, then it is readily shown that
\begin{equation}
{\cal P}(\eta)\propto \eta^\beta,
\end{equation}
and the maximum of ${\cal P}_{\rm tot}$ occurs at
\begin{equation}
\eta_{\rm max} = \eta_0\left(\beta\over \gamma-\beta\right)^{-1/\gamma}\ {\rm or}\ \ 
\eta_{\rm max} = \beta^{1/2}\eta_0,
\label{eq-gauss}
\end{equation}
respectively. 

Knowledge of $\eta_0$ and $\gamma$ would then allow us to limit the
maximum allowed $\beta_i$ for our candidate \physt.  If $W$ has a
power law cutoff, we must have $\gamma > \beta$, or else the
probability would be dominated by $\eta \gg \eta_0$ and we would
discard the \physt\ leading to that $\beta$. The Gaussian falloff in
the conditionalization factor will defeat any power law indices
$\beta_i$, but if $\beta\gg0$, then probability would be peaked at
$\eta \gg \eta_{\rm obs}$ unless $\eta_0 \ll \eta_{\rm obs}$. For
example, if $\beta=$2, 4 or 8, then 95\% of the probability in ${\cal
P(\beta)}{W(\beta)}$ lies at $\eta > \eta_0\times $0.31, 1.04 and 1.8,
respectively.  Thus for our observations to be compatible with
$\beta=8$ at 95\% confidence, we would have to (uncomfortably) assert
that $\eta_0 < \eta_{\rm obs}/1.8$, {\ie} that we are well into the
exponentially cutoff in $W$, and, for example, that a universe with
only slightly more dark matter would have far fewer observers.

An upper limit on allowed $\beta$ can be used in two ways.  First, we
can derive an upper limit to the number components that should, by the
previous arguments, have comparable $\eta_i$, since these components
must all have $\beta_i \ge 1$.\footnote{Of course, if $0\le\beta_i\le
1$, there could be more components; but if the power law extends to
$\eta \ll \edmo$ then the probability lies mostly there, so there is
no reason to expect $\eta\sim\edmo$.}  No such limit exists in the
top-down approach.  Second, a limit on $\beta$ can rule out one
component $i$ having a $\beta_i$ orders of magnitude greater than all
the rest, strengthening (as compared to the top-down approach) the
argument that components with rising $W$ should be comparable in
density -- though as noted above the anthropic argument for multiple
components holds only if $W(\eta)$ is decreasing near $\edmo$.

\section{Equal representation in other cosmological quantities}
\label{sec-per}

The arguments given concerning dark matter that lead to the Principle
of Equal Representation (PER) could also be applied to other
cosmological parameters such as the neutrino density parameter
$\Omega_\nu$, the amplitude $Q$ of density inhomogeneities, and the
cosmological constant $\Lambda$.  In all cases we could place a
constraint on the total value (either from observation or from
anthropic considerations), so we might apply the PER to a situation in
which there were multiple possible contributions to $\Omega_\nu$, $Q$,
or $\Lambda$.  The argument is, however, somewhat different for each
parameter.

In the case of $Q$, inflation tends to quash rival pre-inflationary
perturbation sources; thus a second significant component would have
to be imprinted later, and would likely be either non-Gaussian,
non-adiabatic, or non-scale-invariant (as we presently know no
post-inflation perturbation source without at least one of these
properties).  The present data suggests none of these.

As for $\Lambda$, the argument is complicated by the fact that
$\Lambda$ can be either positive or negative, so that contributions of
large magnitude may cancel. Also, to be meaningfully distinguishable
the contributors must be dynamically different, so the arguments would
have to be applied to multiple parameters describing the details of
the dark energy components.

Massive neutrinos provide an interesting context in which to make
detailed calculations because it is known that there are three
neutrino species with nonzero mass, and because the cosmological
effects of massive neutrinos are relatively well-understood.

Neutrinos with masses in the eV range slow the growth of density
perturbations on small-scales, thereby suppressing galaxy
formation. This effect as been
used~\cite{Tegmark:2003ug,Pogosian:2004hd} to derive an anthropic
prediction of $\Omega_\nu$ (or equivalently the sum $m_\nu=\sum_i m_i$
of the neutrino masses $m_i$) under the assumptions that 1) either
only $m_\nu$, or only $m_\nu$ and $\Lambda$ vary across the ensemble,
2) the number of observers is proportional to the collapse fraction of
dark matter into bound halos, and 3) the prior distribution of either
$m_\nu$ or of the individual neutrino species' masses are flat:
\[
{\cal P}(m_\nu) \propto m_\nu\quad{\rm or}\quad{\cal P}(m_i)\propto m_i.
\] The suppression of galaxies provides a near-exponential cutoff in
$W(m_\nu)$ that when combined with ${\cal P}(m_\nu)$ gives 95\% of the
probability in the range 0.13--5\,eV. (When variation in $\Lambda$ is
included and marginalized over, this shifts to 0.035--5\,eV.)

What would we expect for the neutrino masses using our different
methods of reasoning?  The two measured neutrino mass splittings are
$\delta m_{23}^2=2.5\times 10^{-3}\,{\rm eV}^2$ and $\delta
m_{12}^2=8\times 10^{-5}\,{\rm eV}^2$ \cite{Bahcall03,King03}, and
cosmological observations bound $m_\nu \alt
0.42\,$eV~\cite{Seljak:2004xh}.  From the bottom-up approach, we would
require a \physt\ with high probability to have all three masses $m_i
\alt 0.13$\,eV, and with one family of mass $> 0.007\,$eV and another
of mass $> 0.05\,$eV.  This would be surprising if the families had
independent physics (as in the posited dark matter species discussed
above) but of course they probably do not, and we would simply expect
a successful \physt\ to predict an overall neutrino mass scale $\sim
0.01-0.05\,$eV, with splittings of the same order.

In the top-down approach, the predictions would be similar, except
that it would be possible for the overall neutrino mass scale to be
highly improbable.

The anthropic approach is potentially more interesting.  Under the
assumption that all three neutrino species were governed by
independent ${\cal P}(m_i)\propto m_i$, the anthropic prediction (as
derived by~\cite{Tegmark:2003ug} under the above assumptions) would be
$m_\nu \sim 3\,$eV. The probability distribution for the either of the
two mass splittings can be similarly computed, and is centered about
zero with a width of $\sim 1\,$eV, so we would expect all three
neutrino masses to be $\sim $eV in a nice exhibit of the PER.  (For
the same ${\cal P}(m_i)\propto m_i$, bottom-up reasoning would predict
very large $m_\nu$ and large mass splittings.)  However, As $m_\nu
\sim 3\,$eV violates the cosmological bound and $\sim 1\,$eV
splittings far exceed the observed ones, we must rule out any \physt\
with ${\cal P}(m_i)\propto m_i$.  Instead, we must assume ${\cal
P}(m_\nu)\propto m_\nu$, in which case the predictions
of~\cite{Pogosian:2004hd} are compatible at $95\%$ confidence with the
observed bounds. In this case the mass splittings would presumably
follow from the neutrino physics, rather than the PER.

\section{Variations in multiple parameters, and cosmic coincidences}
\label{sec-coinc}

One of the surprising aspects of the standard cosmological model is
that so many energetic constituents of the present-day universe have
such similar densities, {\ie} $\Omega_\Lambda/\Omega_{\rm DM}\approx
2$, $\Omega_{\rm DM}/\Omega_{\rm b}\approx 5$, and
$\Omega_b/\Omega_\nu\alt 30$, even though many or all of these
densities are thought to arise from essentially independent physics.

Anthropic (or other partial-conditionalization) reasoning affords an
opportunity to explain these ``coincidences" because even if ${\cal
P}(\alpha_i)$ factorizes into ${\cal P}(\alpha_1)...{\cal
P}(\alpha_N)$, the conditionalization factor $W(\alpha_i)$ almost
certainly will not, and therefore the product ${\cal P}W$ used in
making predictions from the theory \physt\ will have correlations
between the $\alpha_i$.  Part of the purpose of the present paper is
to point out that if the true explanation of the coincidences in these
parameters is that their values are bound together by the necessity of
observers, (which provide a particular $W$) then we may expect new,
as-yet-unobserved coincidences. These might consist of similar
contributions to currently unresolvable components such as dark
matter, or, in principle, of new coincidental aspects of the universe.
The ``preposterous" universe~\cite{Carroll:2001xs} may get even worse.

But the possibility of explaining cosmic coincidences comes at a
price: degeneracies in $W$ imply that the anthropic prediction for one
parameter will likely change when additional parameters are allowed to
vary; and there is every reason to believe that degeneracies exist in
$W$~\cite{Aguirre:2001,Tegmark:1997in, Graesser:2004ng}.  Thus an
anthropic explanation of the observed $\alpha_i (i=1..N)$ really
requires the computation of both ${\cal P}$ and $W$ over the
$N$-dimensional parameter space; and until all $N$ are considered,
there is no reason to hope that adding an additional parameter will
not spoil the correct prediction.  In the context of the spectrum of
conditionalization, one is working from the top-down toward a fully
anthropic approach; and as the conditionalization factor is loosened,
more and more theories can be ruled out; each ``successful" anthropic
prediction is only provisional.

\section{Summary \& Conclusions}
\label{sec-conc}

While our observable universe is well-desribed by a simple big-bang
FRW cosmological model, attempts to understand that model at a more
fundamental level through inflation, quantum cosmology, and string/M
theory have raised the spectre of {\em multiple} quasi-FRW regions
(``universes'') with different properties. We must then ask: given a
fundamental theory of physics and cosmology \physt, how can we extract
from it predictions for the single universe we can observe?  It has
been variously asserted that in a future measurement, we should
observe the most probable set of predicted properties (the
``bottom-up'' approach), or the most probable set compatible with all
current observations (the ``top-down'' approach), or the most probable
set consistent with the existence of observers (the ``anthropic''
approach).  These correspond to three different implicit questions of
the form: ``Given that {\em X}, what should I observe in future
measurement {\em Y}'', with $X$ being a conditionalization that is
minimal in the bottom-up approach, and maximal in the top-down
approach.  Given a theory that predicts a multiverse, these questions
will have different answers, and hence different implications what we
will observe in a future measurement.  This paper has been a rough
exploration of these different questions and their differing
implications.

There are both problems of principle and great technical challenges in
mathematically describing the ensemble of universes and defining a
measure so that an {\em a priori} probability distribution ${\cal
P}({\alpha_k})$ can be defined for the parameters $\alpha_k$
describing low-energy physics and cosmology.  Does each possible
universe, or each realized universe, or each volume element, or each
baryon receive equal weight?  Making this choice contitutes a
definition of the bottom-up probabilities, and the ambiguity in this
choice may be termed the ``measure problem" \cite{tegmark03}.

To investigate the top-down and anthropic approaches, we have
optimistically assumed that these {\em a priori} probabilities can in
principle be defined and computed, and have made simple assumptions
about the form of ${\cal P}({\alpha_k})$ in order to draw some
exemplary and qualitative conclusions, focusing on the particular
issue of dark matter, in the context of an imagined ensemble in which
there are $N_{\rm DM}$ {\em physically independent} species of dark
matter, with different densities $\edmi$ ($i=1..N_{\rm DM}$) in each
ensemble member.  Among these conclusions are:
\begin{itemize}
\item{Using bottom-up reasoning, we would generically expect to see
one dominant form of dark matter, because it would be unlikely for any
of the other independent dark components to have a comparable density
by chance.}
\item{Using top-down reasoning, we can fix the total dark matter
density $\eta \equiv \sum_i\edmi$ but not the individual
contributions.  But these can be predicted by assuming that we live in
a rather typical universe consistent with the observed $\eta$. If
${\cal P}(\edmi)$ varies smoothly near $\edmi \sim \edm$ for each $i$,
the overall probability will be maximized for all $\eta_i \sim
\eta/N_{\rm DM}$.  The prediction of many dark matter components {\em
of similar density} is directly contrary to the bottom-up prediction,
and is a demonstration of what we call the Principle of Equal
Representation.}
\item{Using anthropic reasoning, we will obtain the same predictions
for the dark matter components as in top-down reasoning, {\em if} the
conditionalization factor (which essentially constitutes a choice of
definition of ``observer'') forces the prediction for $\eta$, as well
as the other cosmological parameters, to accord with the observed
values (and in particular provides a cutoff at $\edm \agt \edmo$).
Moreover, for some types of ``observers'', the conditionalization
factor $W(\alpha_k)$ can be actually estimated, and used to contrain
${\cal P}(\alpha_k)$ by eliminating theories for which
$P(\alpha_k)W(\alpha_k)$ does not peak near the observed values of
$\alpha_k$.  This strengthens the argument for the coincidence between
components, but suggests that the number of coincident dark components
cannot be large.}
\end{itemize}

None of the three methods (or those elsewhere on the spectrum) is
obviously ``correct", and each has serious issues.  The bottom-up
approach suffers from an ambiguity in the measure, and may rule out
the correct \physt.  The top-down approach may allow an incorrect
theory, by simply asserting that we inhabit an exceedingly unlikely
member of the ensemble, while providing no reason why this is the
case.  And the anthropic method, while providing a measure and a
reason for excluding many ensemble members, requires the definition of
an observer -- which is itself quite ambiguous.

Perhaps, we may hope that nature may give us a clue as to which sort
of reasoning we should employ, for a key point of this paper is that
if the anthropic effects are the explanation of the parameter values
-- and coincidences between them -- that we see, then it ought to
predict that new coincidences will be observed in future
observations. If in the next several decades dark matter is resolved
into several equally important components, dark energy is found to be,
say, five independent substances, and several other ``cosmic
coincidences'' are observed, even the most die-hard skeptic might
accede that the anthropic approach may have some validity.  On the
other hand, if we are essentially finished in defining the basic
cosmological constituents, then anthropic reasoning might be able to
explain some of these, but (beyond its arguably successful prediction
of a small but nonzero cosmological constant) would have missed its
chance predict anything not yet observed.

Probably the best we can hope to do is, starting with a \physt, work
our way down the conditionalization spectrum, knowing that any
successful predictions are provisional as the theory could be ruled
out each successive stage of weaker conditionalization.  But we must
keep in mind that if there is a correct \physt, using it to make
successful predictions may really require some conditionalization, and
we will never be able to be completely sure that there is not a
different \physt\ that could require less.  When dealing with
multiverses, the dream of a final theory may be just that.

\ack We thank Michael Douglas, Patrick Fox, Thomas Hertog, and
particularly Alex Vilenkin for helpful comments on the manuscript.

\end{document}